\documentclass[useAMS,usenatbib]{mn2e}
\usepackage{times}
\usepackage{epsfig,rotating,natbib}

\newif\ifAMStwofonts

\def\msun{M$_{\odot}$}
\def\Msun{M$_{\odot}$ }
\def\bi{\begin{itemize}}

\def\ei{\end{itemize}}
\def\be{\begin{equation}}
\def\ee{\end{equation}}
\def\bea{\begin{eqnarray}}
\def\eea{\end{eqnarray}}

\begin{document}

\title[Fallback accretion] 
{Fallback accretion in the aftermath of a compact binary merger} 
\author[Rosswog]{Stephan Rosswog,\\                   
                         School of Engineering and Science,
                         International University Bremen$^\ast$, 
                         Campus Ring 1, Germany;\\
                         $^\ast$ Jacobs University Bremen as of spring 2007}

\date{}

\maketitle

\label{firstpage}

\begin{abstract}
Recent observations of long and short gamma-ray bursts have revealed
a puzzling X-ray activity that in some cases continues for hours
after the burst. It is difficult to reconcile such time scales 
with the viscous time scales that an accretion disk can plausibly
provide. Here I discuss the accretion activity expected from the material
that is launched into eccentric, but gravitationally bound orbits
during a compact binary merger coalescence. From a 
simple analytical model the time scales and accretion luminosities
that result from fallback in the aftermath of a compact binary merger are
derived. For the considered mass range, double neutron star binaries 
are relatively homogeneous in their fallback luminosities. Neutron 
star black hole systems show a larger spread in their fallback behaviour.
While the model is too simple to make predictions about the detailed 
time structure of the fallback, it makes reasonable predictions 
about the gross properties of the fallback. About one hour after the 
coalescence the fallback accretion luminosity can still be as large 
as $\sim 10^{45}$ erg/s, a fraction of which will be transformed into 
X-rays. Large-scale amplitude variations in the X-ray luminosities can 
plausibly be caused by gravitational fragmentation, which for the 
high-eccentricity fallback should occur more easily than in an accretion disk.
\end{abstract}

\section{Introduction}
Recent observations by Swift and Hete II have confirmed the 
long-held suspicion that short GRBs have a different central engine than long 
bursts: they occur at systematically lower redshifts 
\citep{fox05}\footnote{It has recently been suggested \citep{berger06} that
  at least 1/4 of the short bursts could lie at redshifts $z>0.7$ and would
  therefore implie sunstantially larger isotropised energies than was inferred
  for the first set of detected short GRBs with afterglows.}, in 
galaxies with \citep{hjorth05} and without star formation 
\citep{berger05,barthelmy05}, the emitted gamma-ray energy is substantially 
lower compared to long bursts \citep{soderberg06} 
and no supernova explosion seems to accompany the burst 
\citep{bloom06,hjorth05,fox05}. These observations are naturally 
explained by the most popular central engine for short bursts, the 
coalescence of a compact binary system 
\citep{blinnikov84,eichler89,paczynski91}, but alternative models do exist
\citep[e.g.][]{macfadyen05,levan06}. 
The observations, however, have also revealed long-lasting X-ray activity 
after the main gamma-ray emission phase for about half of the bursts 
\citep{burrows05,nousek06,obrien06}.  Several possible explanations 
have been suggested \citep[e.g.][]{king05,zhang06}, among them is the 
fragmentation of the outer parts of an accretion disk due gravitational 
instabilities \citep{perna06}.\\
This late-time X-ray flaring is considered a challenge to most models 
of short GRBs. In this letter, I discuss the X-ray signature from fallback 
material in the aftermath of a compact binary merger. A fraction of the 
debris material is launched into eccentric orbits and, at some later time, 
falls  back towards the central remnant. For this material, the time scales 
are not set by viscous disk time scales, but instead 
by the distribution of eccentricities within the fallback material which 
is a result of the torques that acted during the merger process. This 
naturally leads to time scales that are orders of magnitudes larger
than usual viscous disk time scales.

\section{A simple analytical model}
\begin{figure}
\psfig{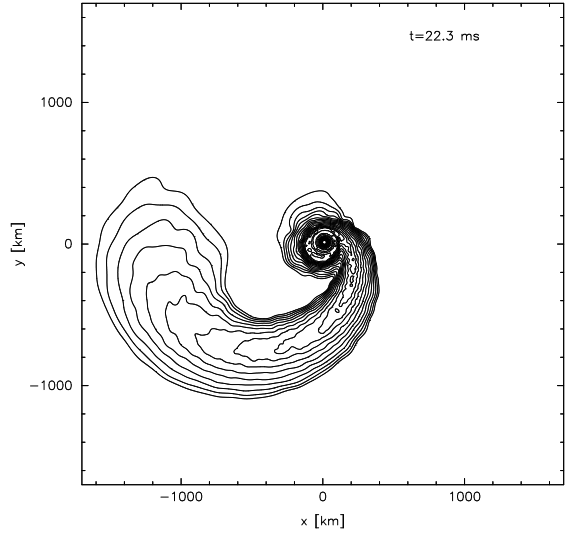}
\caption{Contours of the mass density ($10^7 - 10^{14}$ g cm$^{-3}$) 
  22.3 ms after the merger of a double
  neutron star system with 1.1 and 1.6 \msun. In this case, about 0.03 \Msun
  are launched into eccentric orbits and will finally fall back towards the
  centre.}
\label{fig1}
\end{figure}

The question of fallback time scales is addressed by means of a simple 
analytical model whose initial conditions are taken from  a set of recent, 
three-dimensional smooth particle hydrodynamics (SPH) calculations of the 
merger of both double neutron stars (DNS) and neutron star black hole (NSBH) 
systems. The details of the input physics and the numerical techniques 
are described elsewhere 
\citep{rosswog02a,rosswog03a,rosswog03c,price06b,rosswog06b,rosswog06c}.
Here we only summarise the main physical parameters of the calculations, see
Tab.~\ref{tab:runs}. 
%
\begin{table}
\begin{flushleft}
\caption{Runs that are used as an input for the analytical model.
DNS refers to double neutron star systems, NSBH to neutron star 
black hole binaries. M$_1$, M$_2$ are the masses of the binary 
components (solar units), $q$ is the mass ratio, 
\# part. is the total number of SPH particles 
in the simulation, $m_{\rm fb, -2}$ is the amount of fallback mass in units
of $10^{-2}$ \Msun and $E_{\rm fb}$
the corresponding energy.} 
\label{tab:runs}
\begin{tabular}{rccccccrccccccccc} 
run & type & M$_1$, M$_2$, $q$  & \# part. & $m_{\rm fb, -2}$ &
$\log(E_{\rm fb} [{\rm erg}])$\\ \hline \\
DA  & DNS  & 1.4,  1.4, 1.000     &2 007 516 & 3.61   & 50.9\\
DB  & DNS  & 1.1,  1.6, 0.733     &  400 904 & 3.21   & 51.0\\
DC  & DNS  & 1.1,  1.1,  1.000    &  211 296 & 3.22   & 50.8\\
DD  & DNS  & 1.2,  1.2,  1.000    &  211 296 & 2.84   & 50.7\\
DE  & DNS  & 1.3,  1.3,  1.000    &  211 296 & 3.47   & 50.8\\
DF  & DNS  & 1.5,  1.5,  1.000    &  211 296 & 3.36   & 50.9\\
DG  & DNS  & 1.6,  1.6,  1.000    &  211 296 & 3.61   & 50.9\\
\\
NA  & NSBH & 1.4,  4.0, 0.350    &  600 581 &  7.40   & 51.9\\
NB  & NSBH & 1.4,  6.0, 0.233    &  208 165 &  4.57   & 51.8\\
NC  & NSBH & 1.4, 14.0, 0.100    &2 971 627 &  4.05   & 51.7\\
ND  & NSBH & 1.4, 16.0,  0.088   &1 005 401 &  2.91   & 51.6\\
NE  & NSBH & 1.4, 18.0,  0.078   &1 497 453 &  0.015  & 49.6
\end{tabular}
\end{flushleft}
\end{table} 
%
A few tens of milliseconds after such a binary encounter
the remnant consists of the following
parts: a central object (either a neutron star-like object or a black hole), 
a disk, some material flung into eccentric orbits, but gravitationally 
bound to the remnant (``fallback material'') and some dynamically ejected 
debris material. The details of the geometry depend on the type of system 
(DNS or NSBH) and on the mass ratio. As an example, Fig.~\ref{fig1} shows 
the matter distribution 22.3 ms after the merger of a 1.1 and 1.6 \Msun 
DNS system.\\
Restricted by the Courant-Friedrichs-Lewy time step criterion (typical time
steps are of the order $10^{-7}$ s) the evolution of the merger remnant
can never be calculated hydrodynamically up to the interesting time scales of
hours. We therefore adopt the following simple analytical model.
The SPH-particles that are launched into eccentric orbits, but are still 
bound to the remnant, are treated as test masses in the gravitational 
field of the enclosed mass, $M$, i.e. it is assumed that hydrodynamic 
pressure forces can be neglected and the dynamics of each particle can be
approximated as a point-mass two-body problem. For each of these particles the 
angular momentum, $J_i$, and their energy, $E_i$, can be determined. 
Together with the particle mass $m_i \ll M$ these two quantities yield 
the orbital eccentricity

\be
e_{i}= \sqrt{1+ \frac{2 E_{i} J_{i}^{2}}{G^{2} m_{i}^{3} M^{2}}}. \label{ecc}
\ee
The semi-major axis 

\be
a_{i}= -\frac{G M m_{i}}{2 E_{i}}
\ee
then provides the maximum and the minimum distance of the particle 
orbit from the origin

\be
r_{{\rm max},i}= a_{i}(1+e_{i}), \quad \quad r_{min,i}= a_{i}(1-e_{i}).
\ee
The time until a particle reaches a given radius $R_{\rm dis}$ can be 
calculated by integrating the radial equation

\be
\frac{dr}{dt}= \pm \sqrt{\frac{2}{m_{i}}\left(E_{i}-V(r)-\frac{J_{i}^{2}}{2 m_{i} r^{2}}\right)},
\ee
where $V$ is the potential energy and the positive (negative) 
sign refers to the motion away from (towards) the origin. The time that
elapses while a particle moves from a radius $r_1$ to a radius $r_2$ can
be written as

\begin{equation}
\tau_{r_1,r_2}= \pm \int_{r_1}^{r_2} \frac{r \; dr}{\sqrt{A r^{2}+ B r + C}},
\label{tau}
\end{equation}
where the constants $A= \frac{2 E_{i}}{m_{i}}, \; B= 2 G M$ and 
$C=-\frac{J_{i}^{2}}{m_{i}^{2}}$ have been introduced. The integral appearing
in eq.~(\ref{tau}) can be solved analytically

\be
I_{r_1,r_2}=
\left[ \frac{\sqrt{A r^{2}+ B r + C}}{A} + \frac{B}{2A
    \sqrt{-A}} \arcsin \left( \frac{2 A r + B}{\sqrt{-D}} \right)
\right]_{r_1}^{r_2},
\label{I_ri_R}
\ee
where $D = 4 A C - B^{2}$.
The time that elapses from a particle's present radius, $r_i$, until its 
energy is dissipated at a radius $R_{\rm dis}$ is then given by the contributions 
with the appropriate signs:

\be
\tau_{i}=
\left\{\begin{array}{cl}
I_{r_i,r_{{\rm max},i}} + I_{r_{{\rm max},i},R_{\rm dis}}  
\quad {\rm for} \quad \vec{v}_i\cdot\vec{r}_i > 0\\
I_{r_i,R_{\rm dis}} \hspace*{2.5cm} {\rm for} \quad 
\vec{v}_i\cdot\vec{r}_i < 0
\end{array}\right. \label{fallback_time}.
\ee

\begin{figure}
 \psfig{file=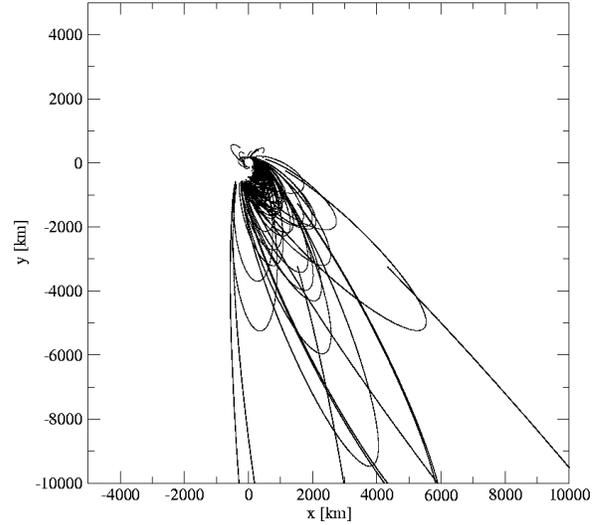,width=8cm,angle=-90}
\caption{A sample of fallback trajectories resulting from a double neutron
  star merger with 1.1 and 1.6 \msun. The dissipation radius, $R_{\rm dis}$ in
  this case was 100 km.}
\label{fig2}
\end{figure}

\begin{figure*}
\psfig{file=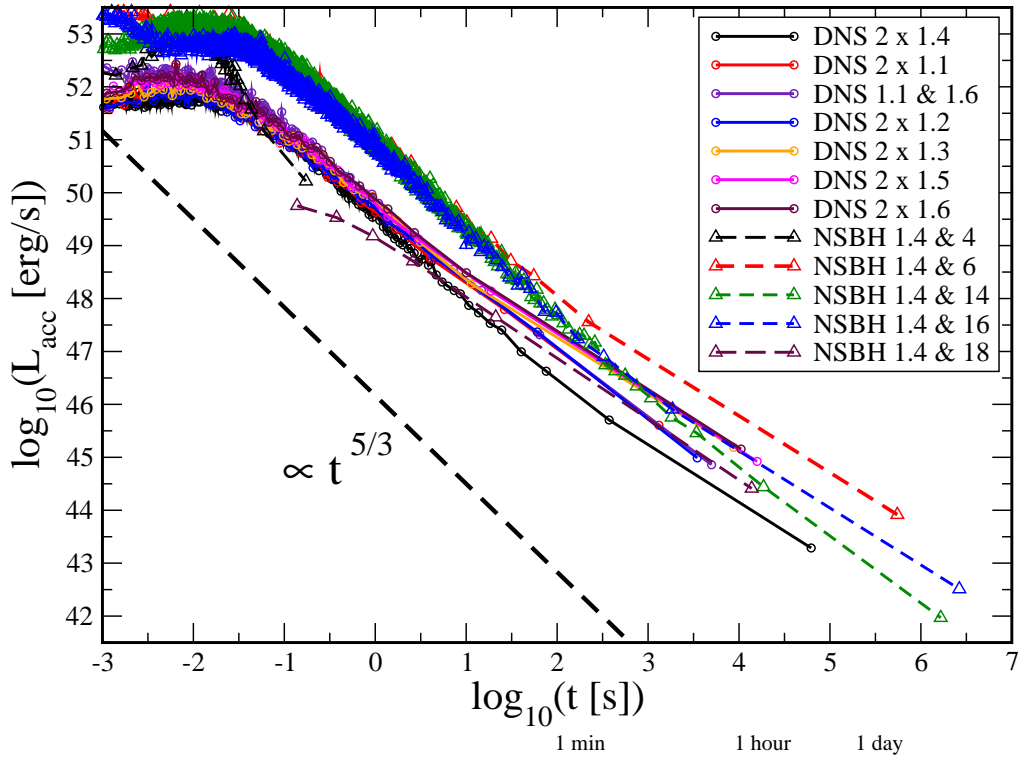,width=13cm,angle=-90} 
\caption{Fallback accretion luminosity for various compact binaries.
Circles refer to double neutron star systems (DNS), triangles to neutron star
black hole mergers (NSBH). Note that the rightmost point is determined by the 
mass resolution of the simulation. For reference a straight line with slope
5/3 is shown.}
\label{fig3}
\end{figure*}
For the radius, where the fallback energy is dissipated, $R_{\rm dis}$, 
we take in the double neutron star case the disk radius at the end of the 
numerical simulation. This is conservative in the sense that the disk will
shrink, see below, and therefore the assumption of a radius fixed at 
$R_{\rm dis}$ yields shorter timescales and lower accretion luminosities. 
For the black hole cases, I choose $R_{\rm dis}= 10 GM/c^2$, so just outside 
the innermost stable circular orbit of a non-rotating (Schwarzschild-) black
hole at $R_{\rm ISCO}= 6GM/c^2$. The details of the fallback times and energies
change slightly with $R_{\rm dis}$, but none of the conclusions of this paper
depends on the exact numerical value of $R_{\rm dis}$.

As an illustration, I show in Fig.~\ref{fig2} a set of fallback trajectories.
The initial conditions were taken from the double neutron star merger
calculation with 1.1 and 1.6 \Msun that is shown in Fig.~\ref{fig2}.
200 randomly chosen trajectories out of the more than 5800 fallback particles
are plotted. There is a broad distribution of eccentricities and fallback
times, while some particles return from their initial position directly 
towards the central object, others follow highly eccentricity orbits out 
to distances of many 10 000 km before they come to a halt and
fall back towards the centre.

\section{Fallback accretion}

Fig.~\ref{fig3} shows the accretion luminosities, 
$L_{\rm acc}= dE_{\rm fb}/dt$, derived for the various DNS and NSBH systems.
Here, $E_{\rm fb}$ denotes the difference between the potential
plus kinetic energy at the start radius, $r_i$, and the potential energy at
the dissipation radius, $R_{\rm dis}$. The curves have been obtained 
by binning the energies contained in the fallback material, $E_{\rm fb}$,  
according to the corresponding fallback times, $\tau_{i}$, see 
eq.~(\ref{fallback_time}). A fraction $\epsilon_{\rm X}$ of this 
energy is channelled into X-rays, $L_X = \epsilon_{\rm X} L_{\rm acc}$.\\
The double neutron star cases form a rather homogeneous class with respect
to their fallback accretion, in all cases the fallback material is
approximately 0.03 \msun, see Tab.~\ref{tab:runs}. After an initial, 
short-lived plateau, the luminosity
falls off with time close to the expected 5/3-power law 
\citep{rees88,phinney89}. It has to be pointed out that
the last point in these curves is determined by the numerical mass 
resolution in the hydrodynamics simulations and should therefore be 
interpreted with some caution. All other points should be a fair 
representation of the overall fallback activity. Typically, the X-ray 
luminosity about one hour after the coalescence is 
$L_X  \sim \left(\frac{\epsilon_{\rm X}}{0.1}\right) \cdot 10^{44}$ erg/s.
For the investigated mass range, the spread in the luminosities one hour after
the coalescence is about one order of magnitude.\\
The neutron star black hole cases show a larger diversity. The mass in the
fallback material of different mass ratios varies by about a factor of 500,
see Tab.~\ref{tab:runs}, an hour after the merger the accretion 
luminosities of the different NSBH systems differ by about two orders of 
magnitude. Also the involved time scales change strongly with the binary mass
ratio. For example, the 1.4 and 4 \Msun NSBH case does not produce much
eccentric fallback material. Accretion, at least to the resolvable level, is
over in $\sim 0.2$ s. This accretion period may produce a short GRB, but
probably not much X-ray activity. The 1.4 and 18 \Msun NSBH system is at the
other extreme: its peak luminosity is lower by three orders orders of
magnitude but extends (at a resolvable level) up to about one hour. The mass
ratios inbetween could possibly produce a (weak) GRB and extended X-ray
activity up to about one day after the burst.


\section{Summary}

A simple analytical model was presented and used to derive the fallback 
time scales
and accretion luminosities in the aftermath of a compact binary merger. The
initial conditions are taken from a large set of 3D  hydrodynamical
simulations. The main assumption of this model is that the fallback 
material is ballistic, i.e. that hydrodynamical pressure forces can be 
neglected. This is a good approximation for the largest part of the time spent
along the individual trajectories, therefore, the fallback time 
scales can be considered valid estimates. The assumption may only break 
down when the vicinity of remnant is approached, but this will only cause 
minor corrections to the fallback time scale and luminosity estimates.\\
For the range of mass ratios (1.1 to 1.6 \msun) that I have considered, 
double neutron star systems show a rather homogeneous fallback behaviour.
At one hour after the coalescence their X-ray luminosity is still 
$\sim 10^{44} (\epsilon_{\rm X}/0.1)$ erg/s. Neutron star black hole mergers 
provide a larger spread in their fallback behaviour. An hour after the tidal
disruption of the neutron star, their range in fallback luminosities is 
about two orders of magnitudes. Some mass ratios could produce a GRB, but
not much X-ray activity, others may produce X-ray flaring only but no GRB,
and yet other systems can produce both.\\
Since this simple approach does not account for self-gravity and gas pressures,
it tends to produce too smooth results. The shown luminosity curves should
therefore be interpreted as temporal averages.
The original accretion disks, such as the one seen in 
Fig.~\ref{fig1} extending out to $\sim 200$ km, are consumed on a 
viscous time scale of 
\be
\tau_{\rm visc} \sim 1/\alpha \Omega_{\rm K} \sim 0.05 {\rm s} \left(
  \frac{R}{200 \;{\rm km}}\right)^{3/2} \left( \frac{0.1}{\alpha}\right)
 \left(\frac{2.5 {\rm M}_\odot}{M_{\rm c.o.}} \right),
\ee
where $\alpha$ is the Shakura-Sunyaev viscosity parameter \citep{shakura73}, 
$R$ the disk radius, $\Omega_{\rm K}$ the Kepler frequency and $M_{\rm c.o.}$
is the mass of the central object. The outer parts of this disk are
constantly fed by material that continuously falls back so that a low 
mass remnant disk is maintained until late times. The observed X-ray 
activity is expected to be produced by this disk and its interaction with 
the high-eccentricity fallback material.\\
Large-scale amplitude variations in the lightcurves (``flares'') can
occur if the fallback material becomes subject to self-gravitational
instabilities. Whether a lump of gas is prone to this instability 
depends on the ratio of the cooling time, $\tau_{\rm cool}$, and 
the local dynamical time, $\tau_{\rm dyn}$. \citet{gammie01} finds
for the case of thin accretion disks the condition 
$\tau_{\rm cool} \le 3 \tau_{\rm dyn}$. Gravitational fragmentation is 
common in the outer regions of the accretion disks of AGNs 
\citep{shlosman90} and of young stellar objects \citep{adams93}. 
\cite{perna06} have suggested that such a fragmentation in the outskirts
of an accretion disk could be the common mechanism behind the flaring 
observed in both long and short bursts. They note, however, that it may be
difficult to reach the suitable conditions in the case of a compact binary
merger. The fragmentation idea has recently been examined in more detail
by \citet{piro06} for the case of collapsar disks. They argue that 
fragmentation may occur near the radius where alpha-particles are 
photodissociated since this consumes about 7 MeV of thermal energy per nucleon
and thus provides an efficient cooling mechanism. They suggest that even 
neutron star-like objects of low masses may form, but conclude that their 
disruption would be too rapid to account for the observed, post-GRB X-ray 
flaring.\\
Some the eccentric fallback material could undergo gravitational 
fragmentation into one or several bound objects. Since there is
a large spread in orbital eccentricities, it seems unlikely that 
a large fraction of the fallback material agglomerates into a single 
fragment. If fragmentation occurs, it will do so only locally and for 
a small fraction of the material.
A fragment of mass $m_{\rm f}$ will release a flare energy of 
$E_{\rm f} \sim 10^{49} {\rm erg} \left(\frac{\epsilon_{\rm X}}{0.1} \right)
\left(\frac{M_{\rm c.o.}}{3 \; {\rm M}_\odot}\right)
\left(\frac{m_{\rm f}}{0.003 \; {\rm M}_\odot}\right) 
\left(\frac{200 \; {\rm km}}{R_{\rm dis}}\right)$.  When 
such a fragment of radius $R_{\rm f}$ falls back 
towards the centre, it becomes tidally disrupted at a distance 
$R_{\rm tid}= (M_{\rm c.o.}/m_{\rm f})^{1/3} \cdot R_{\rm f}$, or, 
if the remnant disk radius is larger than $R_{\rm tid}$, the fragment will 
impact with nearly free-fall velocity on that disk.
In both cases, it will trigger violent X-ray flaring activity.\\
In this picture the GRB is produced by the massive accretion disk 
that can form early on in the merger. The GRB duration 
is determined by the viscous time scale on which this disk is consumed.
The late-time X-ray activity, in contrast, is caused by the fallback of 
a smaller amount, about a tenth of the disk mass, of high-eccentricity 
material which is a natural outcome of the merger. The fallback time 
scales are set by the distribution of eccentricities in this material, 
which is a result of the torques that acted during the merger process.

\section*{Acknowledgements}
The calculations in this paper have partly been performed on the JUMP
computer of Forschungszentrum J\"ulich. 



\end{document}